# UPGRADE OF LINAC CONTROL SYSTEM WITH NEW VME CONTROLLERS AT SPRING-8

T. Masuda, T. Ohata, T. Asaka, H. Dewa, T. Fukui, H. Hanaki, N. Hosoda, T. Kobayashi, M. Kodera,
A. Mizuno, S. Nakajima, S. Suzuki, M. Takao, R. Tanaka, T. Taniuchi, Y. Taniuchi,
H. Tomizawa, A. Yamashita, K. Yanagida
SPring-8, Hyogo 679-5198, Japan

*Abstract*

We integrated an injector linac control system to the SPring-8 standard system on September 2000. As a result of this integration, the SPring-8 accelerator complex was controlled by one unified system. Because the linac was continuously running as the electron beam injector not only for the SPring-8 storage ring but also for NewSUBARU, we had to minimize the hardware modification to reduce the time for the development and testing of the new control system. The integration method was almost the same as that of the integration of the booster synchrotron. We report here on the integration of the linac control system with emphasis on the upgrade of the VMEbus controllers and software involving the operating system Solaris 7 as the real-time OS.

## 1 INTRODUCTION

At the beginning of the SPring-8, accelerator control systems for a linac, a booster synchrotron and a storage ring were designed and constructed independently. Since the storage-ring control system had been defined as the SPring-8 standard, an integration of the synchrotron control system was planned at first and finished on January 1999 as already reported [1]. A linac control system remained to be integrated to the standard system. As already shown in the previous work, the integration of the control system enabled seamless operation between the storage ring and the synchrotron, and made the efficient development of software possible.

The linac control system was originally designed in 1991 [2]. A total number of 25 VME systems were controlled by MVME 147SA-1 (Motorola 68030 CPU) boards with OS-9. The user interface software was developed with Motif on HP-UX. A special protocol was developed and used for the communication between VME systems and a control server workstation. Several X-terminals were connected to the server as the operator consoles.

Recently, system maintenance became more complex and the lack of a data logging system made machine diagnostics difficult.

In September 1999, we started to upgrade the present linac system to the new one by replacing a part of hardware and introducing a standard software scheme as described in below. At this time, we also introduced the standard database system for the linac control [3,4].

Because the linac played a role as an injector to the booster synchrotron and the New SUBARU storage ring, it was necessary to avoid downtime due to the integration procedures. We separated the hardware upgrade work into two phases, that is, we replaced the system with minimum modification at the first phase and postponed an overall replacement to the near future. In the first phase, we replaced only CPU boards for the VME systems keeping the rest of the hardware in place. On the other hand, all the control software was newly developed, and the standard TCP/IP protocol was introduced.

We started preparations in September 1999 by setting the start time of the integration for the July 2000 shutdown period, and finally the new linac system started operation on September 2000.

## 2 NEW CONTROLLER

### 2.1 VMEbus CPU Board

As already described [3], we used Hewlett-Packard HP743rt's for the VMEbus controllers of the storage ring and the booster synchrotron [1]. Because the HP743rt model had been discontinued and it was not possible to purchase, we had to select a new CPU board considering the following criteria:

- The board should be designed on PC-base architecture. The processor unit should be Intel-Architecture (IA-32) or IA-32 compatible.
- The board should support widely used operating systems (OS). An OS with open source policy or a free license is preferred.
- Minimum running costs with easy maintenance and long product lifetime.

Additionally, we required that the board should be a VMEbus single-slot (if possible) or at most a double-slot with a storage unit, bootable from a flash disk, no cooling fan attached to the CPU, and less power consumption to avoid overheating. Usually IA-32 based board is designed on PCIbus base but it can be applied to the VMEbus with a PCI-VME bus-bridge chip (Tundra Universe II). By examining candidate boards, we finally chose a Xycom XVME658 CPU board as the VMEbus controller. The features of the board are:

- CPU is an AMD K6-2 333MHz.
- Bootable from an IDE flash disk.

- Single slot but we use one more slot for the storage units such as the flash disk and a floppy drive.
- No CPU-cooling fan.

It was also essential to decide what OS should be used and examine its operability on the board before we came to the board final selection.

## 2.2 Solaris Operating System

As soon as we had chosen the IA-based CPU, we examined candidate operating systems such as, Linux, RT-Linux and Solaris. To port the current software, the compatibility of UNIX function calls is highly required for the OS. The rigid priority control of software processes is also essential, because it determines controllability of the system under the client/server multi-task architecture. A hard real-time feature such as the absolute deterministic operability is not necessary for our accelerator controls. Task switching latency, data transfer speed and interrupt response time due to a bus-bridge chip were studied as well on the basis of candidate CPU boards [5]. In our study, the Solaris 7 and the Linux OS with a kernel 2.2.9 showed good performance.

The Solaris has three scheduling policies, i.e. a system class (SY), a real-time (RT) class and a time-sharing (TS) class. These classes have a scheduling hierarchy structure as RT>SY>TS. Even in the TS class, the fixed priority control is possible and satisfactory. The RT class is available but its scheduling policy is beyond the system class. So system management with the RT-class is difficult. Because when we lose the control of a process in the RT class the system may hang.

Examining the survey and measurements of the various OS features, we finally chose the Solaris 7 as the OS. We used the TS class for the process scheduling.

# 3 SOFTWARE DEVELOPMENT

## 3.1 Methodology

The equipment experts developed device control software (Equipment Manager, EM [3]) running on the VME systems and graphical user interface (GUI) software on the operator consoles. The control group provided the software framework and program templates. The GUI builder tool and language were same as those of already mentioned [3]. The integration work is summarized as follows:

- Rearrangement of operation sequences and list up of equipment signals to make abstract commands (S/V/O/C format [3]) took three man-months.
- Development of the EM (GUI) software took twelve (eight) man-months, respectively.
- Database parameter set-up and making of access functions took one man-month.

The prototyping methodology was efficient for the rapid development of both EM and GUI programs. The device simulation codes were used for the testing and debugging of the EM, and as a result, the program tests using real devices ran very smoothly.

## 3.2 GUI

The man machine interface for the linac control is categorized in two groups of GUI called *operation panel* and *maintenance panel*. The former is used for machine tuning and leads the beam into five beam transport lines. The later is used for the maintenance of individual equipment.

The linac can be controlled using five *operation panels*, i.e. *top panel*, *gun control*, *RF control*, *magnet control*, and *vacuum operation*. The top panel was newly created on this upgrade as shown in Figure 1. The beam route can be switched by using this panel. The switching procedure includes sequences such as degaussing the bending magnets, opening/closing the beam shutters and setting the transport parameters. Because this procedure was rather complicated, the switching operation took a long time and recorded mis-operations in the previous system. But now, this time has been greatly reduced by introducing this panel.

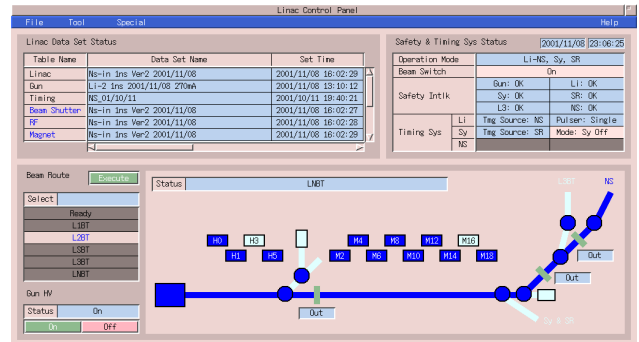

Figure 1: The top panel of the linac control.

As an example of equipment control panels, an RF control panel is shown in Figure 2. The symbols showing machine components are located in the panel based on a real machine layout. Using the panel, the machine operator tunes the beam energy and orbit by monitoring a beam image from profile monitors located along the linac.

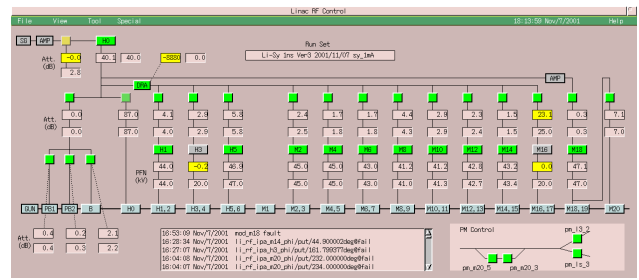

Figure 2: The RF control panel.

As a result of the replacement, the response time between a GUI action and a device reply has been reduced to about 20msec, while it was in the order of a second in the former system. The other significant improvement is that the GUI can be built without knowledge of X/Motif programming. This freedom of development will facilitate an advanced control of the machine.

### 3.3 Database System

The linac has over 270 setting values. A relational database management system consistently manages those values [4]. A total of 84 newly created tables in the database hold a set of parameters. A data acquisition process collects over 5.2kB of data from the linac every five seconds. The database system also keeps this data in 18 tables.

Though the linac control system needed over 100 new tables, the database system minimized the software effort to add them to the database with its standardized procedure.

Now, the linac machine status and logging data can be retrieved from the database, and we can monitor the data via WWW from anywhere in the site. Figure 3 is an example of the machine status monitoring by a WWW browser.

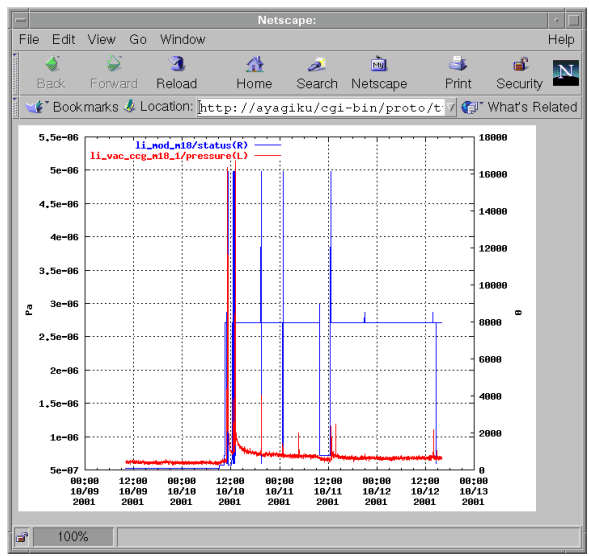

Figure 3: Example of the linac status monitoring with a WWW browser.

## 4 INSTALLATION

We started installation of the new CPU boards in July 2000. After the replacement of the CPU boards, we had some problems apparently caused by electric noise and/or the CPU overheating. The new system was first applied to the aging of the klystrons in the middle of August. During the tuning of the system, the operation of CPU boards stopped either suddenly or relating to the klystron modulator breakdown. Because the VME racks were settled close to the RF modulators, electric noise was the first suspect.

We found that the previous system was not strong enough against the line noise, and also that the cooling system of the VME boards was not well optimized to cool the CPU running at high frequency. After we improved the power line isolation and the cooling system on August 2001, the VME system recorded no trouble for two months.

The Solaris 7 with the TS class has worked well since the installation. It schedules multi-tasks with the rigid priority control and exclusion control at the same level already shown by the HP-RT real-time OS.

## 5 SUMMARIES

We replaced Motorola 68030 CPU boards running OS-9 with new IA-based CPU boards. A set of the XVME658 with Solaris 7 was chosen as the controller for 25 VME systems. We replaced all the control software. The equipment control software on the VME was newly created with the standard framework, and the GUIs for the operator consoles were also developed according to the unified panel look and feel.

We finished the integration of the linac control system to the SPring-8 standard system on September 2000. Now, the SPring-8 whole accelerator complex and beamlines are operated by the unified control system. The new VME system showed a good performance and adequate reliability. The data logging system with the database proved its usefulness again for the machine diagnostics by searching over buried signals.

At this phase, we minimized the hardware modification. However, we are planning to update the VMEbus I/O systems to be better able to withstand against the electric noise and compact enough for easy maintenance in the next phase.